\documentclass[preprint,12pt]{elsarticle}

\journal{Microprocessors and microsystems (MICPRO)}

\usepackage[T1]{fontenc}
\usepackage[utf8]{inputenc}
\usepackage{comment}
\usepackage{graphicx}
\usepackage{todonotes}
\usepackage{caption}
\usepackage{subcaption}
\usepackage[hidelinks]{hyperref}
\usepackage{orcidlink}
\usepackage{paralist}

\usepackage{lipsum}

\begin{document}

\begin{frontmatter}

\title{Cyber-Secure Wearable, Ultra-Low-Power Networked Sensors for Soldier Health Monitoring:
\\
Insights from the CAPSARII project}

\author[sst]{Luciano Bozzi}
\ead{l.bozzi@seaskytech.it}

\author[sst]{Christian Celidonio}
\ead{c.celidonio@seaskytech.it}

\author[sst]{Umberto Nuzzi}
\ead{u.nuzzi@seaskytech.it}

\author[sst]{Massimo Biagini}
\ead{capsarii\_ext\_1@seaskytech.it}

\author[ntnu]{Stefano Cherubin}
\ead{stefano.cherubin@ntnu.no}

\author[ntnu]{Asbjørn Djupdal}
\ead{djupdal@ntnu.no}

\author[ntnu]{Tor Andre Haugdahl}
\ead{tor.a.haugdahl@ntnu.no}

\author[polimi]{Andrea Aliverti}
\ead{andrea.aliverti@polimi.it}

\author[polimi]{Alessandra Angelucci}
\ead{alessandra.angelucci@polimi.it}

\author[polimi]{Giovanni Agosta}
\ead{giovanni.agosta@polimi.it}

\author[polimi]{Gerardo Pelosi}
\ead{gerardo.pelosi@polimi.it}

\author[lwt3]{Paolo Belluco}
\ead{paolo.belluco@lwt3.com}

\author[lwt3]{Samuele Polistina}
\ead{sam.polistina@lwt3.com}

\author[lwt3]{Gaia Strada}
\ead{gaia.strada@lwt3.com}

\author[lwt3]{Luigi Sossi Ortega}
\ead{luigifrancesco.sossi@lwt3.com}

\author[quaesta]{Alexandra-Ioana Albu}
\ead{alexandra.albu@quaesta.ai}

\author[quaesta]{Riccardo Volpi}
\ead{riccardo.volpi@quaesta.ai}

\author[quaesta]{Luigi Malag{\`o}}
\ead{luigi.malago@quaesta.ai}

\author[born]{Michael Schneider}
\ead{michael.schneider@born-germany.de}


\author[ikerlan]{Xabier Eguiluz}
\ead{xeguiluz@ikerlan.es}

\author[ikerlan]{Haritz Saiz}
\ead{hsaiz@ikerlan.es}

\author[ikerlan]{Mar\'{i}a Viqueria}
\ead{mviqueira@ikerlan.es}


\affiliation[sst]{
  organization={Sea Sky Technologies},
  country={Italy}
  }

\affiliation[ntnu]{
  organization={NTNU},
  country={Norway},
  }

\affiliation[polimi]{
  organization={Politecnico di Milano},
  country={Italy}
  }

\affiliation[lwt3]{
  organization={LWT3},
  country={Italy}
  }

\affiliation[quaesta]{
  organization={QUAESTA AI},
  country={Romania}
  }

\affiliation[born]{
  organization={BORN},
  country={Germany}
  }

\affiliation[ikerlan]{
  organization={IKERLAN},
  country={Spain}
  }


\begin{abstract}

The CAPSARII project advances a cyber-secure, ultra-low-power wearable platform and Internet of Battlefield Things (IoBT) framework to support real-time monitoring of military personnel and data-driven decision support. The architecture integrates smart-textile sensing, a body/squad area mesh, edge-deployed AI on secure RISC-V microcontrollers, and cloud/federated analytics for cross-mission learning and comparative studies. Two representative use cases--pilot fatigue and warfighters operating in harsh environments--guide requirements, validation, and interface design (including HUD and haptic feedback).
As an initial step toward embedded models, we benchmark stress classification on a public wearable dataset: tree-ensemble baselines achieve state-of-the-art performance for accuracy and F1 (e.g., XGBoost: $\mathrm{Accuracy}=0.901\!\pm\!0.023$, $\mathrm{F1}=0.920\!\pm\!0.019$), while a feature-based Transformer attains the highest AUC ($0.963\!\pm\!0.025$), informing feature-engineering and compression choices for edge deployment. Building on these results, CAPSARII employs HW/SW co-design, model compression (pruning, quantization, distillation), and precision tuning to meet stringent energy and memory budgets; it further investigates Processing-in-Memory accelerators for future efficiency gains.
A structured gap analysis highlights the remaining challenges: establishing validation standards for physiological signal quality and textile durability; achieving low-latency, scalable BAN/SAN communications under mobility and interference; and deploying hybrid lightweight/post-quantum cryptography with robust device identity and side-channel resilience on constrained endpoints. By consolidating these elements into a coherent, secure IoBT stack, CAPSARII delineates a practical roadmap from laboratory baselines to mission-ready soldier monitoring in both training and operational scenarios.

\end{abstract}



\end{frontmatter}


\section{Introduction}

The increased volatility of the worldwide geopolitical scenario is pushing the European Union to increase its defence capabilities~\cite{andersson2025european}, improving the coordination and cooperation among the member states (and Norway) along 11 priorities identified in the recently revised Capability Development Plan (CDP)~\cite{edf2023cpd}.
The ``Future Soldier Systems'' priority is expected to contribute to, among other goals, improved survivability through the use of wearable sensors for monitoring the physiological and psychological status of soldiers, and the use of Artificial Intelligence to improve the understanding of the collected data.

The European Defense Fund has launched a number of research projects aimed at supporting this goal through the development of new systems.
Among those, the \emph{Cyber-secure weArable, ultra-low-Power networked SensoRs for soldIer health monItoring} (CAPSARII) project aims to develop a novel wearable sensing and computing system designed for use by military personnel, with the specific goal of monitoring fatigue in key operational scenarios.
It is worth noting that wearable devices are widely used in healthcare, sports, and industrial applications, both for studying and enhancing performance and for preventing injuries and accidents~\cite{ometov2021survey}.
However, the defence sector imposes a widely different set of requirements, both in terms of functionality and of non-functional characteristics, which require a dedicated study and design.

Furthermore, while consumer devices rely on civilian standards for interconnection and communication, the CAPSARII system is designed to work within the concept of the Internet of Battlefield Things (IoBT)~\cite{russell2018internet}, which aims to adopt edge computing in scenarios where the network availability can be compromised, to deliver effective and timely information to commanders at the tactical scale, thus increasing tactical awareness and reducing response times.

The CAPSARII wearable system is designed to transform the monitoring of psychophysical conditions and the evaluation of human performance in both combat and training scenarios, providing access to sensor capabilities to monitor a wide range of physiological and environmental parameters, with tactical decision support at the edge tailored to limit the cognitive overhead, while at the same time ensuring the ability to collect, use and share the data for post-mission analysis.
Furthermore, the IoBT system ensures the security and confidentiality of the collected information, and the design takes care to satisfy user needs in terms of battery lifetime and smart textile integration for wearability and durability.

This article extends our earlier overview of the CAPSARII approach~\cite{capsarii-conf}, adding an empirical stress-classification benchmark on a public wearable dataset (Section~\ref{sec:baselineres}) and a structured analysis of the scientific gaps between the state of the art and the project objectives (Section~\ref{sec:gaps}).
We provide an overview of the main goals, subsystems and technological advances needed to reach these ambitious objectives.

The CAPSARII Consortium is composed of small and medium enterprises (SMEs) and research and technology organisations (RTOs) across four EU member countries and Norway. SeaSkyTechnologies leads the consortium, leveraging its expertise as an IoBT system integrator and consultancy company in the defence sector. LWT3 brings expertise in wearable sensors and data analytics, while Quaesta.AI and Born complete the supply chain on the AI and Machine Learning and smart textiles, respectively.
Three RTOs, Politecnico di Milano, the Norwegian University of Science and Technology, and Ikerlan, complete the consortium providing research capabilities across the fields of low-power edge computing, cybersecurity, and bio-engineering.

\section{Use Case Scenarios in Air and Land Domains}

\begin{figure}
	\begin{subfigure}{0.49\textwidth}
	\includegraphics[width=\linewidth]{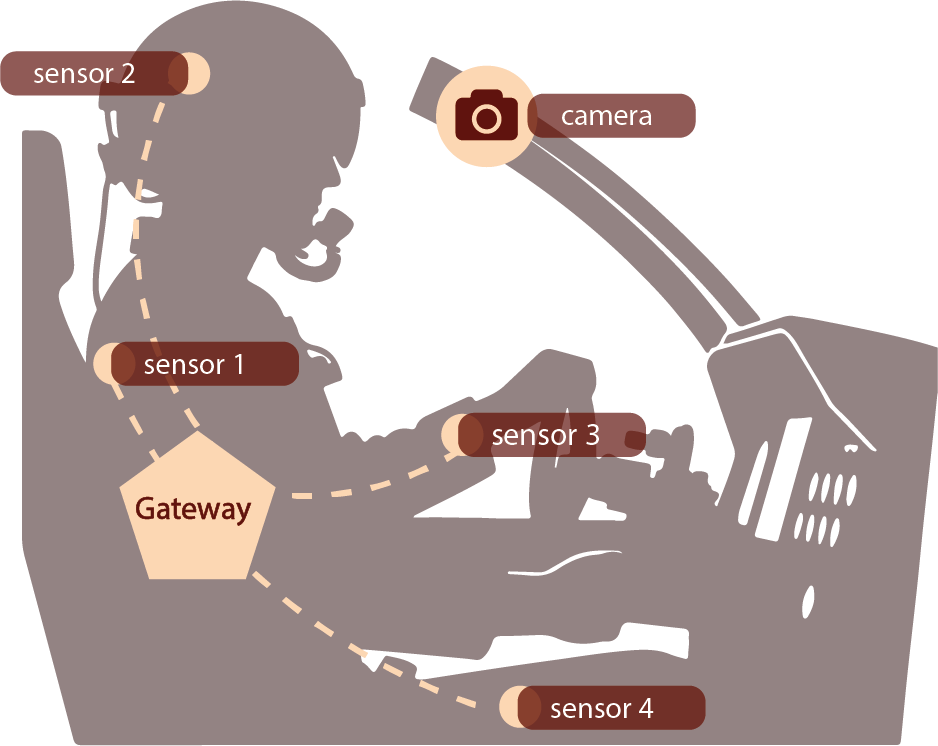}
	\caption{\label{fig:uc1}Use Case 1: Pilot Fatigue}
	\end{subfigure}
	\quad
	\begin{subfigure}{0.5\textwidth}
	\includegraphics[width=\linewidth]{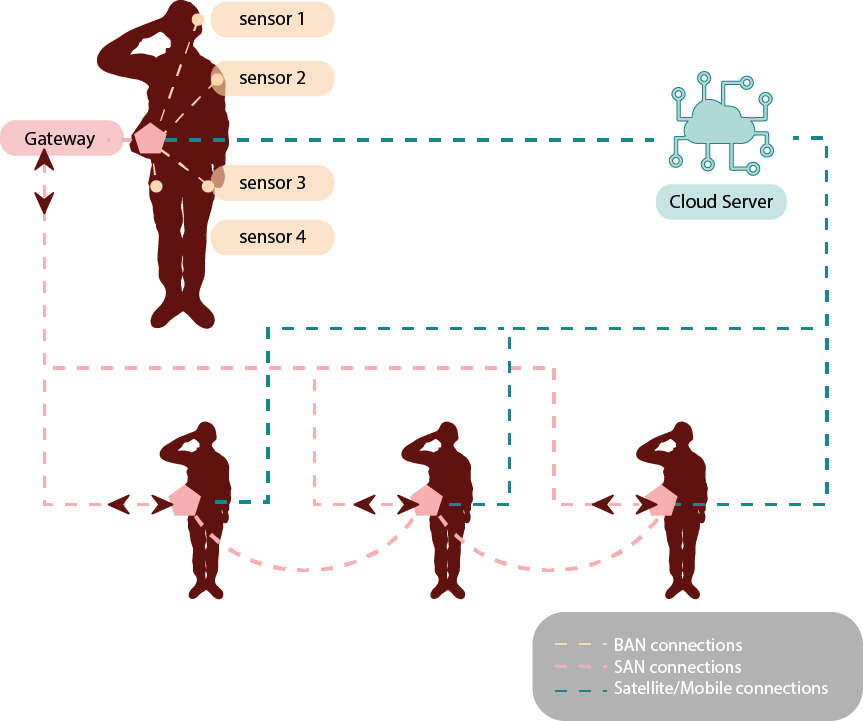}
	\caption{\label{fig:uc2}Use Case 2: Warfighters in Harsh Environments}
	\end{subfigure}
	\caption{\label{fig:uc}CAPSARII Use Cases}
\end{figure}

The proposed CAPSARII wearable system is a template that can be tailored for different use case scenarios.
Within the scope of the project, we consider two of the most important ones, namely \emph{pilot fatigue} and \emph{warfighters in harsh environments}, aiming to demonstrate the versatility of the concept and its ability to address significantly different operational conditions.
On one hand, these two scenarios, depicted in Figure~\ref{fig:uc}, address different domains (air vs land), with different requirements in terms of wearability, mission length, and the nature of fatigue experienced.
On the other hand, both scenarios allow the evaluation of advanced AI features, such as federated learning, that could be useful to address the cooperation needs of multi-force operations, such as joint exercises among EU countries.
In these scenarios, data remain private to the individual force or country, but shared models can be trained in a distributed way.

\paragraph{Pilot Fatigue}
The pilot fatigue use case, shown in Figure~\ref{fig:uc1}, focuses on continuously monitoring the physiological and cognitive state of military pilots during demanding, long-duration missions, where fatigue, stress, and cognitive overload can critically impair performance and safety~\cite{gregory2010pilot,samel2004sleep}.
Alertness and the ability to perform are related to two basic neurophysiological forces: the body’s circadian pacemaker (or biologic clock) and the drive or need for sleep (based on the length of previous wakefulness).
Some countermeasures have been identified, e.g., exposure to cold air, avoidance of night/morning work, naps, caffeine, melatonin~\cite{caldwell2001impact}.
Biomathemathical models can provide personalized fatigue assessment over time~\cite{crs2022,authority2014biomathematical}, which could be used both as a complement to current practices where mission-readiness is assessed through self-evaluation and commander judgement, or to improve sleep management strategies and personalised fatigue countermeasures.
These models integrate multi-modal sensor data to identify early signs of reduced vigilance or micro-drowsiness.
The use case operates under strict constraints of cockpit environments, including minimal acceptable latency for feedback and non-intrusive wearable integration. 

\paragraph{Warfighters in Harsh Environments}
The warfighter in harsh environments use case, shown in Figure~\ref{fig:uc2}, addresses continuous monitoring and support of soldiers operating in highly dynamic, physically demanding, and potentially hazardous conditions such as extreme temperatures, high humidity, or CBRN-threat environments. 
Real-Time Physiological Status Monitors (RT-PSM)~\cite{friedl2016real} are an emerging class of modern military wearable technologies, providing individual soldiers and their leadership with actionable physiological status information needed to ensure individual and squad health and performance/readiness.
The CAPSARII system equips the soldier with smart-textile-based wearables integrating multi-modal sensors to capture physiological signals (e.g., heart rate, respiration, muscle activity, stress via GSR), motion patterns, and environmental parameters including temperature, humidity, gas exposure, and air quality. The system is designed to operate reliably under severe constraints (e.g., -15 to $\mathrm{+55^\circ C}$, high shock/vibration, and limited connectivity), ensuring data integrity and robustness. Real-time processing is enabled through edge AI, which fuses heterogeneous sensor streams with millisecond synchronization to detect fatigue, stress, or dangerous conditions and trigger timely alerts. 
Communication relies on resilient Body Area Networks (BAN)~\cite{telfer2020open} and Squad Area Networks (SAN)~\cite{lai2009digital}, maintaining situational awareness across units even in degraded or contested environments. 
Energy efficiency is critical, with ultra-low-power design supporting continuous multi-day missions without recharge. Modular sensing and interoperability with portable CBRN detection devices further extend operational capabilities. Overall, this use case demonstrates how CAPSARII enhances soldier safety, endurance, and mission effectiveness by providing real-time health monitoring, environmental awareness, and AI-driven decision support in extreme operational scenarios.


\section{State-of-the-Art and Available Technologies}

\begin{figure*}
\begin{center}
	\includegraphics[width=\textwidth]{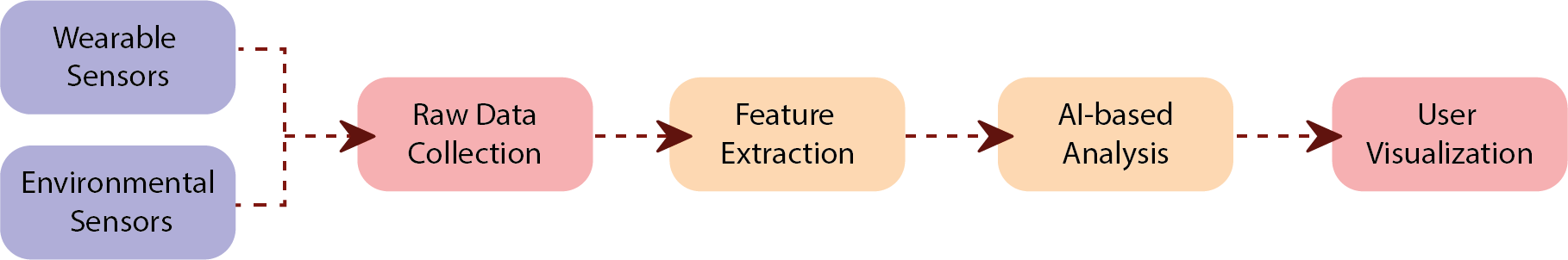}
	\caption{\label{fig:pipe}The CAPSARII data collection and analysis architecture}
\end{center}
\end{figure*}

The CAPSARII project is founded on the concept of Internet of Battlefield Things (IoBT), which interconnects military-grade equipment into a protected IoT computing architecture.
This system, depicted in Figure~\ref{fig:pipe} is envisioned as a continuous, interconnected process, ensuring efficient data flow and analysis to and from the wearable devices.

\paragraph{Data collection to the edge}
At the lowest tier, the wearable devices continuously acquire physiological, movement, and environmental signals, together with operational metrics, and forward them over a mesh network~\cite{capsarii-conf}. Rather than shipping every raw sample upstream, an early stage of computation is carried out close to the point of acquisition---on the wearable itself or on a nearby edge node. Performing this first evaluation locally keeps the latency of time-critical inferences low and reduces the volume of data that must traverse the network before an actionable decision can be reached.

\paragraph{From the edge to the server}
The edge tier does not attempt to be exhaustive: once a preliminary analysis has been produced, the data deemed relevant---whether summarised features or selected raw segments---is forwarded to a central server for deeper processing and long-term storage.

\paragraph{Cloud-based analysis}
At the server tier, the CAPSARII IoBT system draws on cloud-based analytics for computationally heavier evaluation that the edge cannot sustain. A distinctive aspect of the design is its collaborative dimension: analytical capacity can be pooled across cloud resources belonging to different EU nations, enabling assessments, cross-subject correlation, and pattern mining at a scale unattainable on a single node or deployment.

\paragraph{Visualisation}
The outcome after processing is surfaced to authorised personnel through a visualisation layer that supports timely, informed decisions. A particular focus is the delivery of feedback directly to the operator: CAPSARII investigates Head-Up Displays (HUDs) and other media channels that combine audiovisual and tactile stimuli to communicate critical information under a tight cognitive budget. A HUD can, for instance, render vital signs such as heart rate, respiratory rate, and stress level, alongside environmental factors such as temperature, humidity, and pollutant concentration, using compact icons or colour coding. In parallel, haptic cues such as vibration---modulated in frequency and amplitude---offer an unobtrusive feedback path in situations where visual or audio channels are unavailable or undesirable.

In combination, these tiers integrate wearable sensing, layered data analytics, operator-centric interfaces, and federated cloud strategies into a single IoBT stack aimed at the demands of military operations while fostering cooperation across European defence. The remainder of this section surveys the state of the art and available technologies underlying its principal building blocks: smart textiles and wearable system integration, the IoT architecture, the body area network, security, AI-model optimisation, and the emerging class of ultra-low-power AI processors.

\subsection{Smart Textiles}
Smart textiles are fabrics into which electronic components are embedded so that the garment itself can sense, actuate, communicate, and process data. 
The field has matured rapidly, with progress on flexible and stretchable sensors and actuators, on-textile energy harvesting and storage, and increasingly miniaturised integrated electronics.
As textiles begin to process the signals they acquire rather than merely capture them, they open a transformative design space that increases device sophistication, connectivity, durability, safety, and energy efficiency.
For a military deployment, coupling the textile to the wider IoBT, addressing user training, and weighing ethical considerations are equally central. Ergonomics calls for a multidimensional treatment of comfort for military personnel \cite{knight2002comfort}, while sustainability motivates recycling and reusing materials to support a circular economy, and the practical integration of these garments into training and operational routines aims to improve safety and effectiveness.

\subsubsection{Wearable System Integration}
Integration of advanced wearable devices with soldiers' gear, capturing real-time data on physiological, movement and environmental parameters, and operational metrics, by studying the integration and embedding of state-of-the-art sensors into wearable devices. Particular focus will be put into eye-tracking with an event-based camera that can perform similarly to high-end commercial detectors.
Furthermore, the wearable devices provide soldiers and commanding officers with intuitive interfaces for data submission, visualisation, and two-way communication, focusing on reducing cognitive overload.
Parametric design principles (specific for operational theatres and roles), allow for the customization of wearable devices, data analytics, and visualisation interfaces.
Integration with the soldier's gear needs to take into account that the devices will be compliant with necessary safety and reliability standards (e.g. STANAG and ATEX directives), such as the standards for use in potentially explosive environments, like aircraft refuelling stations and hangars.

\subsection{IoT Architecture}

An IoT architecture is a framework defining the structure and components of an IoT system, describing how devices and services interact to collect, process, and exchange data through secure mechanisms, in compliance with data-sovereignty regulations. It is typically organised into several layers, each with a specific function \cite{zhong2017study}.
The \emph{perception layer} hosts the sensors and actuators, which in CAPSARII are embedded in the wearable devices. The \emph{network layer} transports the data gathered by the perception layer toward the central processing unit,
performing some basic preprocessing such as aggregation, abstraction, and analysis. The \emph{application layer} carries out the actual data processing, delivering insight, monitoring, and basic automation.
The IoT applications and user interfaces also sit in this layer, letting users access and interact with the system through web applications, mobile apps, or other clients. 
For the IoBT system, CAPSARII will design an IoT architecture whose device layer consists of wearables on a mesh network with dynamic master nodes, an edge node with AI capabilities for low-latency decisions, and a server layer running on different cloud platforms.
This architecture should support an efficient flow of data from the wearables to external servers, scale on demand, and include redundancy for system resilience. It should combine edge computing for initial data processing, secure transmission protocols, and cloud-based federated analytics for in-depth evaluation and pattern mining, reducing latency and improving real-time decision making through early on-device analysis. 

\subsection{Body Area Network}
A Body Area Network (BAN), sometimes referred to as a Body Sensor Network (BSN), interconnects the small, low-power sensors worn on or close to the body and relays their readings, over short-range wireless links, to a nearby aggregator such as a smartphone or a fixed monitor. Among the protocols suited to this setting, Thread~\cite{unwala2018thread} is a prominent open standard for low-power wireless IoT devices. Layered on top of IEEE~802.15.4, it establishes a reliable, secured mesh and was conceived for home automation and similar deployments in which low power draw, scalability, and robustness are paramount. Its self-healing mesh recovers automatically when individual nodes drop out, and it further provides native IPv6 addressing, built-in security, and interoperability---properties that are attractive for battery-operated endpoints. Where lower latency and higher throughput are required, Zenoh~\cite{zenoh} can be layered above Thread.
Within CAPSARII, the BAN is a target both for hardening and for optimisation. On the security side, we investigate mechanisms that strengthen authentication, encryption, and message integrity so as to blunt cyberattacks directed at edge devices and their networks. On the performance side, the objective is to cut transmission latency and raise throughput without eroding---and ideally while improving---energy efficiency, device authentication speed, data-integrity guarantees, resistance to security breaches, and overall network uptime.

\subsection{Security}

\subsubsection{Device Identity and Authentication}

Managing the identities of end devices is a multifaceted problem, and one that grows sharper as the number of connected devices increases. An identity-management system must scale to that population while preserving secure authentication and, at the same time, respecting user privacy. Because such systems concentrate trust, they are themselves attractive targets, so their own robustness is a first-order concern. Interoperability across devices that adhere to different identity standards adds a further layer of complexity, as does the need to govern the full identity lifecycle---from provisioning through decommissioning---a stage that is frequently neglected.
Striking the right balance among these pressures is what ultimately safeguards the integrity and confidentiality of digital identities.
Within CAPSARII, the aim is to design an efficient identity-management platform tailored to the project's end devices, accounting for their heterogeneity and resource constraints while meeting the elevated security and privacy expectations of military settings. As part of this effort, we also consider the emerging threat that quantum computing poses to current identity and key-exchange mechanisms~\cite{nist-pq}.

Within CAPSARII, security enforcement is centralized at the endpoint wearable device, which acts as the system trust anchor. Flexible and textile-integrated sensors are treated as untrusted or minimally trusted data sources and do not perform cryptographic operations, secure storage, or identity management.

\subsubsection{Endpoint Security for RISC-V}

Security for the RISC-V ecosystem is attracting growing attention, as witnessed by the work of the RISC-V International Security Standing Committee on cryptographic extensions and trusted execution environments~\cite{lu2021survey}. That work has so far centred on well-established primitives such as AES and SHA-2, whereas the ultra-low-power regime targeted by CAPSARII calls for lighter alternatives; here, recent standardisation competitions~\cite{caesar,nist-lc} have surfaced strong candidates for lightweight yet secure encryption.
CAPSARII concentrates its endpoint-security work at the microcontroller, the boundary at which the platform meets the outside world. The specific research direction is the interplay between lightweight cryptography, dedicated hardware acceleration, and protection against side channels and covert channels~\cite{agosta2020compiler}.

\subsection{AI Models}

\subsubsection{Signal acquisition and pre-processing pipeline}\label{features}

As a preliminary analysis we can establish a benchmark, utilizing the publicly available stress detection framework and dataset originally developed by Hongn et al. \cite{hongn2024wearable}. The dataset contains physiological signals recorded with an Empatica E4 device for 36 subjects during three sessions that included stress-inducing or physical activities. The stress inducing sessions included a Stroop Test, a version of the Trier Mental Challenge Test, stating the personal opinion on a controversial topic, followed by stating the opposite opinion and a subtraction test. The tests were interleaved with resting periods.  The measurements of three of the subjects were discarded due to incorrect placement of the device or connectivity issues \cite{hongn2024wearable}. For creating the dataset, we followed the procedure described by Hongn et al., in which signals were segmented and split into stressful activities and rest periods, thus obtaining a binary classification task. In our experiments, two data representations were investigated: one in which the raw signal was used for training and one in which several features were extracted from the signal. The set of features was derived similarly to Hongn et al., from  heart rate information, accelerometer data, blood volume pulse and electrodermal activity. A first subset of features includes statistics computed on the heart rate signal: the mean and standard deviation of the original, unfiltered signal and the minimum and maximum change ratios, given by the 5\% and 95\% percentiles of the first derivative of the signal. The accelerometer data was used to extract the following features: the mean and standard deviation computed for each of the three axes, as well as the mean, standard deviation and the minimum and maximum change ratios of the magnitude. The Blood Volume Pulse (BVP) data was first filtered using a 4th-order Chebyshev Type II digital filter and the mean and standard deviation of this filtered signal across time were obtained. The BVP data was also used to derive heart rate variability (HRV) information. Systolic peaks were identified and both time and frequency features were extracted using the NeuroKit2 \cite{Makowski2021neurokit} library. The time features capture different statistics related to the intervals between consecutive peaks (normal RR intervals): the standard deviation of the intervals, the square root of the average squared differences between successive RR intervals, the percentage of absolute differences computed for consecutive intervals which are greater than 20/50 ms, the minimum, maximum and mean of the RR intervals and the average heart rate. The frequency features measure the distribution of power across different frequency bands: high frequencies (HF) and very high frequencies (VHF). Additionally, the total spectral power (TP) and the normalized HF (HF/TP) were computed.
The last subset of features was computed using Electrodermal Activity (EDA). The EDA signal was first filtered using a 5th-order Butterworth low-pass filter and then decomposed into its tonic and phasic components. The mean and standard deviation of the raw EDA signal were added to the feature set. In addition, the mean, standard deviation, minimum and maximum change ratios were computed for the tonic component. For the phasic component, the mean and standard deviation were calculated. Additionally, several features were obtained using Neurokit2: the number of peaks per signal length, the mean amplitude of the peaks, their mean height, the mean recovery time and the mean rise time.
The obtained features were normalized using z-score normalization.

\subsubsection{Selected baseline AI models}


Several AI models can be analyzed on the stress classification task, namely: an eXtreme Gradient Boosting (XGBoost) \cite{chen2016xgboost} model, a Random Forest (RF) \cite{breiman2001random}, a Support Vector Machine (SVM) \cite{hearst1998support}, a fully connected neural network and a transformer architecture \cite{vaswani2017attention}.

The XGBoost and RF are both tree-based models. The RF is an ensemble of decision trees which uses a type of bagging \cite{breiman2001random}, while XGBoost trains multiple decision trees through boosting \cite{chen2016xgboost}. The optimal hyperparameters for XGBoost were selected  by varying the maximum depth of the trees (from 3 up to 6) and the step size shrinkage (\texttt{eta} $\in \{0.1, 0.2, 0.3 \}$). The optimal RF model was found by investigating different number of trees (100, 200, 300).

SVMs are distance-based classifiers which build a hyper-plane separating the instances belonging to different classes. In order to be applicable for datasets with instances that are not linearly separable, SVMs use the so-called kernel trick \cite{hearst1998support}. We experimented with a regularization hyper-parameter $C\in \{1, 5, 10\}$, polynomial and RBF kernels and kernel coefficient $gamma \in \{0.001, 0.01, 0.1 \}$.

In the experiments using XGBoost, RF and SVM, the signals are represented using the feature set described in section \ref{features}. Additionally, we trained a fully connected network on this data representation. The network architecture consisted of 5 fully connected layers followed by batch normalization and the ReLU activation function. The network was trained for 100 epochs using the AdamW optimizer and a learning rate of 0.001.

Additionally, a Transformer model was studied for the stress classification task. The Transformer architecture \cite{vaswani2017attention} has been successfully used across a wide range of tasks, from natural language processing to computer vision. A central component of the architecture is the multi-head self-attention module, which allows capturing relationships between different tokens in a sequence without using recurrent neural networks.
We represented the physiological signals as sequences of partially overlapping 30s windows. Embeddings were obtained for the windows in two ways: using the features described in section \ref{features} and using a convolutional feature extractor applied on the raw signals. The model using the features extracted from the signal used 8 attention heads, while for the model using the convolutional feature extractor the optimal number of attention heads was 16. The best results were obtained using learned positional encodings. The architecture contained a single transformer encoder layer with a dimension of 128 neurons for the internal linear layer. A classification token was added to the beginning of the sequences and its embedding was used as input to the classification layer. The models were trained for at most 200 epochs, using the AdamW optimizer and a learning rate of 0.001. The training of the model was stopped after 5 epochs with no decrease of the validation loss.

\subsubsection{Evaluation metrics and test methodology}

The performance of our models was evaluated using several metrics: Accuracy, Precision, Recall, $F_1$ score and Area Under the Receiver Operating Characteristic Curve (AUC). 80\% of the subjects were used for training, 10\% for validation and the remaining 10\% for testing. The splitting was repeated five times and the mean and standard deviation of the obtained evaluation metrics were reported.

\subsubsection{Baseline performance summary}
\label{sec:baselineres}

Table \ref{results} shows the results obtained by evaluating the ML models on the stress classification task. The best results in terms of Accuracy, Precision Recall and $F_1$ score were obtained by the XGBoost model, while the best AUC was obtained by the Transformer using the feature representation.
When comparing the two transformer models, it can be observed that the Transformer trained on features extracted from the windows outperformed the Transformer trained on the raw signal, which highlights the relevance of the extracted feature set.

\begin{table}[h]
\resizebox{\textwidth}{!}{
\begin{tabular}{|l|l|l|l|l|l|}
\hline
Model                  & Accuracy         & Precision & Recall & F1     & AUC    \\ \hline
SVM                    & 0.861$\pm$0.055         & 0.852$\pm$0.046    & 0.934$\pm$0.072 & 0.890$\pm$0.046 & 0.940$\pm$0.032 \\ \hline
RF                     & 0.868$\pm$0.060           & 0.844$\pm$0.069    & 0.967$\pm$0.031 & 0.900$\pm$0.043 & 0.962$\pm$0.018 \\ \hline
XGBoost                & \textbf{0.901$\pm$0.023}          & \textbf{0.905$\pm$0.024 }   & \textbf{0.935$\pm$0.023 } & \textbf{0.920$\pm$0.019 } & 0.948$\pm$0.031 \\ \hline
Fully connected        & 0.803$\pm$0.052 & 0.862$\pm$0.037    & 0.806$\pm$0.113 & 0.829$\pm$0.056 & 0.884$\pm$0.046 \\ \hline
Transformer (raw)      & 0.856$\pm$0.058           & 0.889$\pm$0.073    & 0.881$\pm$0.107 & 0.880$\pm$0.050 & 0.962$\pm$0.026 \\ \hline
Transformer (features) & 0.875$\pm$0.054           & \textbf{0.905$\pm$0.055}    & 0.889$\pm$0.080 & 0.895$\pm$0.046 & \textbf{0.963$\pm$0.025} \\ \hline

\end{tabular}

}\caption{Performance of the compared models on the stress classification task (mean $\pm$ standard deviation). Best results are shown in \textbf{bold.}}\label{results}
\end{table}



\subsubsection{Optimisation for Efficiency}

The rapid expansion of artificial intelligence has catalyzed advancements in image classification and object detection \cite{pouyanfar2018survey}. However, the increasing architectural complexity of deep learning models presents substantial deployment challenges for resource-constrained hardware, such as smartphones and IoT sensors \cite{li2018deep}, which operate under far more stringent memory and power limitations than centralized data centers \cite{Chen2019Deep}. While cloud-based inference offers superior processing power, it introduces critical bottlenecks regarding latency and data privacy.
Within the CAPSARII framework, we address these constraints by employing a multifaceted model compression strategy to optimize AI architectures for wearable systems. This includes pruning to eliminate redundant connections \cite{han2015deep}, quantization to reduce the precision of weights and activations \cite{courbariaux2015binaryconnect}, and knowledge distillation to train compact student models that emulate high-capacity teacher networks \cite{hinton2015distilling}. Together with weight sharing \cite{ullrich2017soft}, these techniques facilitate the deployment of robust, low-power AI models at the edge.


\subsection{Ultra-Low-Power Platforms for Edge AI}
The CAPSARII use cases impose signifiant requirements on the system's energy efficiency, in particular for the warfighters in harsh environments scenarios, where extended missions running for over 48 hours strain the ability to the computing platform to sustain continuous multi-modal sensing, AI processing, and communication without recharging. 
To this end, a careful combination of hardware and software strategies is necessary, according to the well-established paradigms of hardware/software codesign and platform-based design of embedded systems~\cite{demicheli2013hardware}.
This methodology is applied in two directions. On one hand, aiming at a rapid prototyping, we select appropriate off-the-shelf computing platforms that can support the computation and communication needs of the wearable system.
On the other hand, we explore future-oriented solutions based on a secure RISC-V microcontroller as the common platform element, with a range of accelerators, from Processing-In-Memory (PIM) to embedded GPU-like accelerators, to compose the right solution for specific applications.

\subsubsection{Processing-in-Memory}
Since many AI-based applications are heavily memory-bound, due to the need to process and aggregate large amounts of data with relatively simple operators, PIM solutions have been proposed to move computation in or near memory, either by attaching compute elements with limited capabilities near the memory (e.g., with 3D stacking, or by modifying SRAM arrays to perform parallel operations), or by performing analog computations by exploiting the physical properties of the chips~\cite{asifuzzaman2023survey}.
Specifically, near-memory computing refers to DRAM-based solutions leveraging 3D stacking or compute units embedded in a DIMM\cite{kwon202125,devaux2019true}, while in-memory computing includes both digital solutions (usually SRAM-based) and analog solutions (based on non-volatile memory crossbar arrays employed to perform vector-matrix multiplications)~\cite{ielmini2018memory}.
%
CAPSARII explores the use of PIM technologies to speed up and improve the energy efficiency of the execution of its ML tasks.

\subsubsection{Precision Tuning}
Since many AI-based applications exhibit significant degrees of error-tolerance, they are a good target for trading off computation accuracy for performance.
Such trade-offs can be exploited by reducing the size of the data (e.g., moving from double precision to single precision IEEE754 floating point types) or by switching from floating point to fixed point~\cite{cherubin2020survey}.
In CAPSARII, we leverage the developments in precision tuning tools carried out in previous research efforts in the civilian sector, such as  H2020 FETHPC ANTAREX~\cite{8344645} and the EuroHPC TEXTAROSSA~\cite{agosta2022textarossa} projects, by extending the Tuning Assistant for Floating-Point to Fixed-Point Optimization (TAFFO) framework~\cite{Cattaneo2022TAFFO} to support new architectural solutions, such as embedded GPUs and PIM accelerators, thus contributing to the roadmap of mixed precision computing~\cite{fornaciari2023hardware}.
In CAPSARII, we also aim at a deeper characterization of performance improvements based on the characteristics of the target architecture and application~\cite{le2018mixed,denisov2024impact}.


\section{Scientific gaps}
\label{sec:gaps}

This section identifies the technical and scientific gaps between the state-of-the-art and the goals of the CAPSARII project.
This gap analysis is focused around the two CAPSARII use cases: Warfighters in harsh environments and Pilot fatigue.
This is because many of the gaps are relevant only for one use case and not the other one.
Sections~\ref{subsec:gap-harsh-env} and \ref{subsec:gap-pilot-fat} lists all identified gaps for these use cases.
There are, however, many gaps in the state of the art that are relevant for both use cases, or not related to any of the use cases but still relevant for CAPSARII.
Those gaps are listed in Section~\ref{subsec:gap-others}.


\subsection{Warfighters in harsh environments}
\label{subsec:gap-harsh-env}

\paragraph{Physiological Status Monitoring}



There is a lack of suitable validation standards for the signal quality in physiological sensors.  The CAPSARII project will develop validation standards to maintain consistent signal quality--such as acceptable signal-to-noise ratios and effective motion artefact suppression--following repeated mechanical stresses and laundering procedures.

\paragraph{Wearable Feedback Technologies}


Current HMDs can render overlay data and various information streams, but they are unable to display the full range of data envisaged by the CAPSARII project.
A further critical dimension concerns ergonomics and usability. Parameters such as weight, comfort and system performance continue to be decisive factors in determining the operational acceptability of these devices among soldiers. State-of-the-art solutions currently only partially mitigate these limitations.
%
%
Current military platforms are not yet capable of managing the full range of biometric data and other data types from individual soldiers and entire battalions. Aggregating such data will require the development of a visualization platform that should be structured in a shared and hierarchical manner and consistent with existing practices for other categories of military information.

\paragraph{Smart Textiles}

For ground soldiers, smart textiles aim to deliver enhanced protection against environmental stressors (cold, heat, moisture), alongside improved mobility and comfort. Recent advancements include battery-free, chipless textile interfaces capable of wireless power and data transmission, reducing hardware bulk and improving washability and flexibility \cite{revaiah2019technical, xu2025}. Existing challenges include poor durability, the possibility of incorporating multiple functions in a single smart textile, and field validation of the smart textiles.


\subsection{Pilot fatigue}
\label{subsec:gap-pilot-fat}

\paragraph{Physiological Status Monitoring}


The existing suite of physiological parameters is not fully optimised for monitoring individuals in high-G or hypoxic environments. It is necessary to assess SpO2 tracking, respiration morphology, and thoracic effort as supplementary metrics alongside conventional heart rate and respiratory rate measurements.

Cockpit environments present unique challenges, including variations in cabin pressure and the potential accumulation of CO2 or volatile organic compounds (VOCs). Data fusion strategies should be developed to correlate these environmental exposures with indicators of cognitive fatigue.


\paragraph{Wearable Feedback Technologies}

HMDs are primarily designed to convey environmental information.  For example, they do not display pilot fatigue levels directly, which would be useful for preventing injury and muscoskeletal disease.
As the interface becomes more complex, it will be necessary to redesign how the pilot can access and visualize these new information streams, including their placement within the display and the frequency with which they are presented.

Currently, pilot vital health parameters are not managed in an integrated way. These data should be made available for sharing at all levels of the chain of command during the mission. Furthermore, current systems do not monitor pilots' physiological parameters prior to mission execution.




\paragraph{Smart Textiles}

For military pilots, smart textiles are vital for mitigating extreme stresses such as high G-forces, temperature fluctuations, and fire risk. While developments like flame-retardant blends and thermal adaptive materials are advancing, integration remains limited.  Reliability in extreme conditions must be improved, and the trade-off of comfort vs. protection is still unresolved.  Finally, integration of smart textile solutions with flight gear is not fully standardized.


\subsection{Additional gaps}
\label{subsec:gap-others}

\paragraph{Physiological Status Monitoring}



Analyzing the literature, it emerges that most research to date has focused on monitoring individual parameters and comparing them with selected endpoints, such as perceived physical and/or cognitive fatigue. However, multimodal approaches to physiological status monitoring remain underdeveloped, despite their importance: accurate assessment of fatigue requires the simultaneous acquisition and synchronization of multiple signals.

While several wearable devices are already available for monitoring specific physiological parameters, what is still missing is an integrated system capable of combining these devices into a comprehensive, real-time monitoring platform for soldiers.
Finally, a further obstacle to the development of a comprehensive monitoring system lies in the fact that many wearable devices currently available on the market are produced by companies outside the EU.

\paragraph{The Battlefield Computing Continuum}


Many traditional IoT protocols (e.g., ZigBee, Thread) are designed for static mesh networks. However, in applications where nodes often move rapidly, split, and rejoin the network, smooth device discovery and reconnection are essential.  There are alternatives, such as Zenoh and MANET protocols.  Most of the MANET use-cases are, however, not intended for military environments, where excessive signaling may spoil the users location. It is likely that none of the available discovery mechanisms are perfectly suited for CAPSARII and new ones must be developed.


Protocols like MIOTY and LoRaWAN handle thousands of nodes effectively, but their low data rates limit use cases with continuous streaming or complex sensor fusion. High-density networks require scalable protocols that balance throughput and reliability, particularly for soldier SANs with multiple sensors transmitting simultaneously.  No clear candidates were found in the state-of-the-art.



Applications such as health monitoring or node coordination require ultra-low latency communications. BLE and UWB offer low latency for short-range links, but continuous monitoring of multiple sensors, especially under high mobility, may introduce latency spikes. Edge processing, temporary local storage, and protocols like Zenoh, MQTT, or WebSocket could help, but extreme conditions remain challenging.

The continuum integrates heterogeneous devices and networks. Protocols like Matter promote interoperability at the application layer, yet full cross-layer interoperability between physical, network, and application protocols is still evolving.

Most terrestrial IoT protocols, such as WiFi, ZigBee and LoRaWAN, rely on pre-existing infrastructure or proximity to gateways, limiting their effectiveness in remote, rural or disaster-affected areas. While satellite IoT solutions offer global coverage, there are often trade-offs in terms of bandwidth, power consumption and cost. Seamless integration of terrestrial and non-terrestrial networks remains a key challenge for ubiquitous IoT connectivity.
Satellite based star topology could be the only feasible solution for pilots in a squadron, but on the ground more power-efficient solutions can be developed based on MANET protocol stacks, which discover neighboring devices and route data through them. An efficient and reliable network topology awareness and routing optimization are still key challenges for MANETs.




\paragraph{Low Power Wearable Systems}

Precision tuning software is the chosen low-power software technique employed by CAPSARII.
It is currently unknown the extent to which the typical guarantees of target accuracy provided by precision tuning software, such as TAFFO, apply to the domain of AI models.
If Processing-in-Memory (PIM) devices are included in CAPSARII, research and development are needed on precision tuning to enable a tool such as TAFFO to apply energy-efficiency oriented optimisations in a programming paradigm employing PIM.



\paragraph{AI-Based Signal Processing}

Future research will prioritize the integration of these methodologies into a unified, mission-ready wearable solution within the CAPSARII framework. Building upon the feature-based transformer models detailed in Section \ref{sec:baselineres}, subsequent phases will focus on optimizing these architectures for deployment on edge devices, balancing high predictive accuracy with the stringent computational constraints of embedded systems.

A central objective involves addressing data variability through granular dataset segmentation. By categorizing data according to physical activity (e.g., athletes vs. non-athletes), age demographics, and environmental conditions (warm vs. cold testing), the framework provides the necessary dimensionality to enhance meta-learning and federated learning approaches. To mitigate the challenge of reduced sample sizes within these specific cohorts, the implementation of few-shot learning and methods with high inductive bias will be essential to maintain model robustness across all groups.

Regarding neurophysiological monitoring, efforts will concentrate on overcoming the inherent noise and domain-shift challenges associated with event-based sensors in deployed environments. Initial validation will involve a rigorous evaluation of tracking and velocity estimation algorithms using the 27-person Berkeley event-based eye-tracking dataset \cite{angelopoulos_2025}. This systematic parameter tuning and benchmarking phase serves as a critical step in advancing event-based eye-tracking techniques and ensuring the precision of the project's tracking solutions.

\paragraph{The Security Dimension}

Within the framework of IoBT, security remains a pivotal challenge, particularly in projects such as CAPSARII, where wearable, ultra-low power devices must operate in adversarial and resource constrained environments. Although significant progress has been made in the areas of PQC and LWC, several open research gaps hinder their effective integration into battlefield networks.

A first gap relates to the maturity of standardization and adoption in military-grade systems. While the NIST process has established first generation standards such as Kyber (KEM) and Dilithium (digital signatures) in the PQC domain, and ASCON as the reference algorithm for LWC, the uptake of these primitives in distributed, defense-critical architectures remains at a very early stage. Current literature indicates a lack of doctrinal and architectural guidelines on how to embed these primitives into IoBT deployments, which raises questions regarding interoperability, lifecycle management, and compliance with NATO and EU defense standards.

A second challenge concerns the compatibility of PQC with constrained hardware. Post-quantum schemes -- particularly lattice-based constructions -- introduce non-negligible overheads in terms of memory footprint, bandwidth consumption, and computational complexity. This makes their deployment in wearables and textile-embedded sensors problematic, where energy autonomy and form-factor constraints are critical. By contrast, lightweight cryptographic primitives are well aligned with these constraints, but they provide no intrinsic resilience against quantum adversaries. The lack of hybrid cryptographic approaches that can jointly address near-term feasibility (through LWC) and long-term sustainability (through PQC) represents a significant gap in both academic research and applied defense innovation.

A third research frontier lies in the domain of key management and secure scalability under battlefield conditions. IoBT environments are characterized by node mobility, intermittent connectivity, and exposure to physical and cyber attacks. The large key sizes and complex key exchange mechanisms associated with PQC amplify these difficulties, while LWC offers efficiency but lacks standardized frameworks for large-scale, post-quantum-capable key distribution. Research into lightweight post-quantum key management protocols--capable of supporting thousands of mobile nodes in adversarial conditions--remains at an embryonic stage.

Finally, there is limited evidence regarding the operational resilience of PQC and LWC implementations under realistic military conditions. Both families of algorithms require validation not only against conventional cryptanalytic and side-channel attacks, but also in the presence of electronic warfare, jamming, and resource exhaustion strategies typical of contested battlefield environments. The absence of systematic experimental campaigns or large-scale testbeds in this area underscores a major gap between algorithmic research and operational deployment.

In summary, while PQC and LWC have individually advanced considerably in recent years, their joint application within the IoBT paradigm has yet to be comprehensively addressed. The current gap can thus be characterized as the absence of standardized, hybrid, and empirically validated cryptographic protocols that simultaneously ensure quantum resilience and lightweight efficiency. Addressing this convergence is not only an open research question but also a strategic imperative for future European defense systems, including those envisaged by CAPSARII.


\section{Conclusion}

The CAPSARII project aims at developing an innovative IoBT wearable system and related ICT framework for supporting military staff. Tactical decision support is enabled through continuous monitoring of physiological, movement, and environmental parameters, complemented by AI models deployed directly on edge nodes. Two use cases guide the work:
\begin{inparaenum}
\item Pilot fatigue, where the wearable system provides real-time awareness of crew fatigue and suggests personalised countermeasures;
\item Warfighters in harsh environments, where individual and team health status is monitored to prevent stress- and fatigue-related incidents.
\end{inparaenum}

As an initial step toward embedded deployment, we benchmarked stress classification on a public wearable dataset: tree-ensemble models (XGBoost) lead on accuracy and F1, while a feature-based Transformer achieves the highest AUC. These findings inform the feature selection and compression strategies for edge-ready models. A structured gap analysis further identifies three critical R\&D fronts: (i) validation standards for signal quality and textile durability, (ii) robust low-latency BAN/SAN communications under mobility, and (iii) hybrid lightweight/post-quantum cryptography with side-channel resilience on constrained devices. Finally, preliminary results on RISC-V-based PIM accelerators quantify the energy trade-offs of Processing-in-Memory for future efficiency gains.

Through these contributions, CAPSARII consolidates wearable sensing, HW/SW co-design, efficient edge AI, and IoBT security into a coherent stack, and delineates a practical roadmap from laboratory baselines to mission-ready soldier monitoring in both training and operational scenarios.



\section*{Acknowledgement} This work has been funded by the European Union via the European Defence Fund project CAPSARII under grant number 101168102.
Views and opinions expressed are however those of the author(s) only and do not necessarily reflect those of the European Union or the European Defence Fund.
Neither the European Union nor the European Defence Fund can be held responsible for them.

\bibliographystyle{elsarticle-num}
\bibliography{refs}

\end{document}